\documentclass[prl,aps,twocolumn]{revtex4}
\usepackage{graphicx}
\usepackage{amsmath}
\usepackage{bm}

\usepackage{times}

\newcommand{\be}{\begin{equation}}
\newcommand{\ee}{\end{equation}}
\newcommand{\bea}{\begin{eqnarray}}
\newcommand{\eea}{\end{eqnarray}}
\newcommand{\bei}{\begin{itemize}}
\newcommand{\eei}{\end{itemize}}

\begin{document}

\title{Stable Topological Superfluid Phase of Ultracold Polar Fermionic Molecules}

\author{N. R. Cooper$^{1,2}$ and G. V. Shlyapnikov$^{2,3}$}
\affiliation{\mbox{$^{1}$T.C.M. Group, University of Cambridge, Cavendish Laboratory, J.J. Thomson Ave., Cambridge CB3 0HE, UK}\\
\mbox{$^{2}$Laboratoire de Physique Th{\'e}orique et Mod{\`e}les Statistiques,  Universit{\' e} Paris Sud, CNRS, 91405 Orsay, France}\\
\mbox{$^{3}$van der Waals-Zeeman Institute, University of Amsterdam, Valckenierstraat 65/67, 1018 XE Amsterdam, The Netherlands}}

\begin{abstract}

  We show that single-component fermionic polar molecules confined to a 2D
  geometry and dressed by a microwave field, may acquire an attractive $1/r^3$
  dipole-dipole interaction leading to superfluid $p$-wave pairing at
  sufficiently low temperatures even in the BCS regime.  The emerging state is
  the topological $p_x+ip_y$ phase promising for topologically protected
  quantum information processing. The main decay channel is via collisional
  transitions to dressed states with lower energies and is rather slow,
  setting a lifetime of the order of seconds at 2D densities $\sim 10^8$
  cm$^{-2}$.

\end{abstract}
\date{17 July 2009}
\maketitle

Remarkable progress in the studies of ultracold atomic Fermi
gases\cite{Varenna,Trento} has opened up prospects for creating novel
phases of fermionic atoms. Of particular interest is the topological
superfluid $p_x+ip_y$ phase for identical fermions in two dimensions
(2D) \cite{Radzihovsky}, discussed in the
contexts of superfluid $^3$He and the fractional Quantum Hall effect
\cite{Volovik,Green}.  The intense interest arises from the exotic
topological properties of the phase at positive chemical potential
$\mu>0$ (i.e. in the BCS regime).  In the presence of vortices, the
groundstate becomes highly degenerate, spanned by zero-energy Majorana
modes on the vortex cores \cite{Green,sternannals}. The highly
non-local character of these states is expected to suppress
decoherence processes, and allow this degenerate subspace to be used
for topologically protected quantum information processing
\cite{nayakrmp}.

The $p_x+ip_y$ topological phase has been predicted to be the groundstate of
ultracold fermionic atoms interacting via a $p$-wave Feshbach resonance
\cite{Radzihovsky}. However, the realization of the $p_x+ip_y$ phase in this
way encounters serious difficulties. Away from a Feshbach resonance the
superfluid transition temperature is vanishingly low. While it may be
increased on approach to the resonance, in this case the system becomes
collisionally unstable. Fermions form long-lived diatomic quasibound states
and their collisions with the atoms cause relaxation into deep molecular
states, leading to a rapid decay of the gas \cite{Gurarie,Castin}.

In this paper we show that a stable topological $p_x+ip_y$ phase can be
created with fermionic polar molecules with large dipole moment. Ultracold
clouds of polar molecules in the ground ro-vibrational state have been
obtained in recent successful experiments \cite{Jin,Weidemuller}. Fermionic
$^{40}$K$^{87}$Rb molecules \cite{Jin} have a permanent dipole moment $d\simeq
0.6$ D, and the dipole moment of $^6$Li$^{133}$Cs fermionic molecules should
be close to 6 D, the same as for the created bosonic molecules
$^7$Li$^{133}$Cs \cite{Weidemuller}.  Being electrically polarized such
molecules interact via long-range anisotropic dipole-dipole forces, which has
crucial consequences for the nature of quantum degenerate regimes. In
particular, this provides the possibility of superfluid pairing at
sufficiently low temperatures in a single-component Fermi gas. In 3D the
ground state of a gas of fermions with dipole moments aligned in the
$z$-direction has a pairing function which vanishes for $p_z=0$
\cite{Baranov}. The 2D Fermi gas of canted dipoles has a ground state with a
pairing symmetry of a similar form \cite{Bruun}.  In both cases, the presence
of nodes in the order parameter make these phases distinct from the $p_x+ip_y$
topological phase.  Our route to a stable $p_x+ip_y$ phase is somewhat
simpler than other approaches\cite{zhangsatonishida}.

Our idea is to use polar molecules confined to a 2D geometry and
dressed by a microwave (MW) field which is nearly resonant with the
transition between the lowest and the first excited rotational
molecular levels. As we describe below, the dressed polar molecules
acquire an attractive $1/r^3$ dipole-dipole interaction, which leads
to superfluid pairing of $p_x+ip_y$ symmetry. Staying in the BCS
limit, the superfluid transition temperature can be made sufficiently
large, and decay processes sufficiently slow, to allow realization of
this phase in experiment. The effects of MW dressing of polar
molecules have been considered as a way to tune the
intermolecular potential\cite{Micheli}, and to form a repulsive shield
for suppressing inelastic losses\cite{Gorshkov}. The possibility of
attractive interactions, which we consider here, has very important
consequences for the nature of the phases that can arise, and for the
stability.

We consider a gas of fermionic polar molecules which are tightly confined in
one ($z$) direction and assume that the confinement length $l_z$ still greatly
exceeds the size of a molecule.  Then the translational motion of the
molecules is 2D, but rotational eigenstates $|J,M_J\rangle$ are those of the
3D molecule.  The operator of the dipole moment $\hat{\bm{d}}$ can have
non-zero matrix elements only between states with different rotational
quantum numbers $J$. The transition dipole moment for $J=0\,\rightarrow
J=1$ is $d_t\!=\!|\langle0,0|\hat{\bm{d}}|1,M_J\rangle|\!=\!d/\sqrt{3}$, with
$M_J\!=\!0,\pm 1$, and $d$  the permanent dipole moment of the molecule.

We then apply a circularly polarized MW field which propagates in the $z$
direction and has a frequency $\omega$ close to the frequency $\omega_0$ of
the transition between the states $|0,0\rangle$ and $|1,1\rangle$. If the Rabi
frequency $\Omega_R\equiv d_tE/\hbar$ and the detuning
$\delta\equiv \omega-\omega_0$ satisfy the inequality
$|\delta|,\Omega_R\ll\omega_0$, then the rotating wave approximation is valid
and the MW electric field ${\bm E}(t)$ couples only the states $|0,0\rangle$
and $|1,1\rangle$.  The resulting states may be represented in
the dressed-molecule picture, with wavefunctions\cite{gauge}
\begin{eqnarray}
  |+\rangle  & = &a |0,0;N\rangle +b e^{-i\omega t}|1,1;N-1\rangle \label{eq:g1}
\label{eq:a}
 \\
 |-\rangle  &= &b  |0,0;N\rangle -a e^{-i\omega t}|1,1;N-1\rangle,  \label{eq:g}
\label{eq:b}
\end{eqnarray}
where $N$ labels the number of photons in the field, and
$a=-A/\sqrt{A^2+\Omega_R^2}$, $b = \Omega_R/\sqrt{\Omega_R^2+A^2}$,
$A=(\delta +\sqrt{\delta^2+4\Omega_R^2})/2$. We will
consider $\delta\agt \Omega_R$ , and choose $\delta > 0$ such that the
energy of the state $|+\rangle$ lies {\it above} the energies of
$|-\rangle$ and $|1,-1\rangle$.  If the MW field is ramped on
adiabatically, then the ground state $|0,0\rangle$
evolves into the state $|+\rangle$, and all molecules can be prepared
in this state. As described below, relaxation to the
lower energy states, $|-\rangle$ and $|1,-1\rangle$,  can be very slow.

We derive the interaction potential between two molecules within the
Born-Oppenheimer approximation, in which the molecules are assumed to be at
fixed locations with a separation $\bm{r} =r(\cos\phi,\sin\phi)$.  At large
separations, the molecules are both in the state $|+\rangle $.  Each molecule
has an effective electric dipole moment $d_{\rm eff}=-\sqrt{2}a b d_t$, which
rotates in the plane of translational motion: $\langle + | \hat{\bm d} |
+\rangle = d_{\rm eff}(\cos\omega t, \sin\omega t,0)$. The interaction
potential at large distances is then
\begin{equation}      \label{eq:vt} 
\!V(r)\!=\!\frac{\bm{d}_1\bm{d}_2\!-\! 3 (\bm{d}_1\hat{r})(\bm{d}_2\hat{r})}{r^3}\!=\!
\frac{d_{\rm eff}^2}{r^3}\!\left[\!1\!\!-\! 3\cos^2 (\omega t\!-\! \phi)\right]\!.\!\!\!\!\!
\end{equation}
Thus, the {\it time-averaged} interaction is attractive:
\begin{equation}
\label{eq:effectiveinteraction}
V_0(r\to\infty)=-d^2_{\rm eff}/2r^3
\end{equation}
and is characterized by the lengthscale $r^*\equiv Md^2_{\rm
eff}/2\hbar^2$, where $M$ is the mass of a molecule. The quantity $r^*$ is defined
analogously to the van der Waals length for atoms and is a measure of the
radius of the centrifugal barrier experienced by the (fermionic)
molecules.

\begin{figure}[ttp]
\includegraphics[width=0.98\columnwidth]{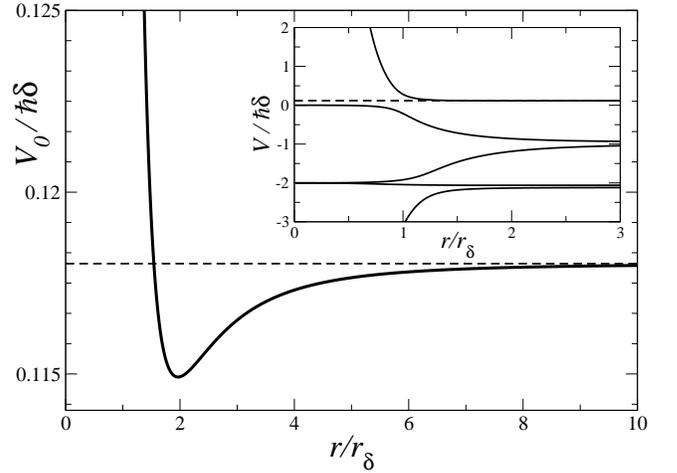}
\caption{ \label{fig:bosurface} Potential energy curve
  $V_0(r)$ for two $|+\rangle$ state molecules, computed for 
  $\Omega_R = 0.25 \delta$ (see text). Anticrossings with other field-dressed levels
  of even parity occur at distances $r\sim r_{\delta}$, as shown in the inset.}
\end{figure}

At smaller separations, the dipolar interactions between the molecules cause
them to depart from the state $|+\rangle$. This occurs when the characteristic
interaction energy $d_t^2/r^3$ becomes larger than the detuning
$\hbar|\delta|$, setting a new lengthscale $r_\delta\equiv
[d_t^2/(\hbar|\delta|)]^{1/3}$.  We have found the resulting Born-Oppenheimer
surfaces using a full coupled channel calculation containing all the levels
$|J,M_J; N\rangle$. Similar calculations are described in
Refs.~\cite{Micheli,Gorshkov}.  We choose a positive detuning $\delta>0$ and assume
that the lengthscale of the potential, $r_\delta$, is larger than the
confinement length $l_z$, so that the interaction is 2D. The potential energy
curves of even parity are illustrated in Fig.\ref{fig:bosurface}, showing a
potential $V_0(r)$ that has a repulsive core for $r\alt r_{\delta}$ and is
attractive at $r\agt r_{\delta}$, with a long-range $1/r^3$ tail.
As discussed below, the repulsive core prevents low-energy
particles from approaching each other at distances $r\alt r_\delta$ and
suppresses inelastic collisions, including 
``ultracold chemical reactions'' recently observed at JILA for KRb molecules.

We now analyze the low-temperature phase of a 2D gas of identical fermions
interacting via the potential $V_0(r)$, assuming that $r_{\delta}\ll r^*$. Due
to the presence of an attractive $1/r^3$ tail given by
Eq.~(\ref{eq:effectiveinteraction}), one expects that the Fermi gas is
unstable to the formation of a superfluid state. In the ultracold dilute
limit, where the momenta of colliding fermionic particles satisfy the
inequality $kr^*\ll1$, this tail provides a contribution $\propto (-kr^*)$ to
the scattering amplitude. This is the so-called anomalous contribution coming
from distances of the order of the de Broglie wavelength of the particles and
obtained in the Born approximation \cite{LL3}. It greatly exceeds the leading
short-range ($r\alt r^*$) contribution which is related to the $p$-wave
scattering and is $\propto k^2$ away from $p$-wave resonances. Thus, omitting
second order corrections, a detailed behaviour of the potential $V_0(r)$ at
distances $r\alt r^*$ drops out and the only important lengthscale is $r^*$.
 
In our analysis of the superfluid phase we confine ourselves to the BCS weak
coupling regime, where $k_Fr^*\ll 1$ with $k_F=\sqrt{4\pi n}$ being the Fermi
momentum and $n$ the gas density. (We consider a uniform 2D gas; effects of a
trap can be included within the local density approximation.)  The regularized
gap equation is obtained expressing the interaction potential through the zero
energy vertex function $\Gamma({\bm k},{\bm  q})$ governed by \cite{AGD} 
\begin{equation}            \label{GammaV}
\Gamma({\bm k},{\bm q})=V_0({\bm k}-{\bm q})-\int\frac{d^2{\bm q}^{\prime}}{(2\pi)^2}\frac{\Gamma({\bm k},{\bm q}^{\prime})V_0({\bm q}-{\bm q}^{\prime})}{E_{q^{\prime}}},
\end{equation}
with $E_q=\hbar^2q^2/2M$, and $V_0({\bm q})$ being the Fourier transform of
$V_0(\bm{r})$. The gap equation 
then reads \cite{Leggett,Randeria2d,Trento}:
\begin{equation}     \label{reggap}
\Delta_{{\bm k}} = -\int \frac{d^2\bm{q}}{(2\pi)^2}\Gamma({\bm k},{\bm
  q})\frac{\Delta_{{\bm q}}}{2}\left[\frac{\tanh(\epsilon_{{\bm
        q}}/2T)}{\epsilon_{{\bm q}}}-\frac{1}{E_q}\right],
\end{equation} 
where $\epsilon_{{\bm q}}=\sqrt{ (E_q -\mu)^2+ |\Delta_{{\bm q}}|^2}$ is the
energy of single-particle excitations, and $\mu>0$ is the chemical potential
which is equal to the Fermi energy $E_F=\hbar^2k_F^2/2M$. To first order we
replace $\Gamma({\bm k},{\bm q})$ in Eq.~(\ref{reggap}) with $V_0({\bm k}-{\bm
  q})$. At $T=0$ we put $\tanh(\epsilon_{{\bm q}}/2T)=1$ and perform an
analytical analysis assuming that in the weak coupling limit the main
contribution to the integral in Eq.~(\ref{reggap}) comes from momenta $q$
close to $k_F$. It shows that the dominant pairing instability is in the
channel with orbital angular momentum $l=1$. The most stable low temperature
phase has $p_x\pm i p_y$ symmetry, following from the fact that this phase
fully gaps the Fermi surface, in contrast to competing phases
\cite{andersonmorel}. A full numerical solution of the regularized gap
equation confirms this analysis. It further shows that $|\Delta_{{\bm k}}|$
rises linearly for $k\alt k_F$, and approaches a constant $\sim
E_F\exp(-3\pi/4k_Fr^*)$ for $k\agt k_F$.

In the 2D geometry that we consider, the critical temperature $T_{\rm c}$ of a Fermi gas is set by the Kosterlitz-Thouless transition.
However, in the weak coupling limit the Kosterlitz-Thouless 
temperature is very close to $T_{\rm c}$ obtained in the BCS approach \cite{miyake}.
For $T\rightarrow T_{\rm c}$ we omit $|\Delta_{{\bm q}}|$ in the expression for
$\epsilon_{{\bm q}}$ in Eq.~(\ref{reggap}) and obtain:
\begin{equation}             \label{eq:critTc}
T_{\rm c}\approx E_F\exp(-3\pi/4k_Fr^*),
\end{equation}
where the numerical prefactor is of order of unity \cite{footnote}.
Thus, to obtain an achievable value of $T_{\rm c}$ one requires
$k_F r^*$ to be not much smaller than unity. The BCS approach assumes that the
exponential factor in Eq.~(\ref{eq:critTc}) is small and $T_{\rm
  c}\ll E_F$.  A limitation on the strength of the attractive
interaction is set by the condition of stability to phase separation (collapse
to a high-density gas). A full calculation of this limit requires a
strong-coupling theory. However, estimates 
(provided by Hartree-Fock theory) suggest that the
compressibility is positive for $k_Fr^* < 3.7$. Thus, the regime of moderately
strong interactions $k_F r^*\sim 1$ is accessible.

The $p_x + ip_y$ phase spontaneously breaks time-reversal invariance (the
phase $p_x - ip_y$ is its degenerate time-reversed partner)\cite{comment1}.
It can be viewed as a state in which the Cooper pairs have an orbital angular
momentum of $\hbar$ with respect to the $z$-axis.  The $p_x+ip_y$ phase
can exist in one of two topologically distinct phases, depending on the sign
of the chemical potential\cite{Volovik}. The phase at $\mu<0$ may be
continuously deformed to the vacuum state; the phase at $\mu>0$ is
topologically distinct from the vacuum and has several very interesting
properties.  Most notably, the vortices of this phase carry localized
zero-energy states, described by a Majorana fermion on each vortex core. These
lead to non-abelian exchange statistics \cite{Green,sternannals} and possible applications for topologically protected quantum information
processing\cite{nayakrmp}. The superfluid of dipolar interacting spinless
fermions that we describe above has $\mu>0$ and is in the relevant
topological phase.

The typical interatomic potential between atoms or molecules (without MW
dressing field) has a short range $R_0\sim 1 - 10$ nm. Thus, in 2D the
scattering phase shift is $\sim (k_FR_0)^2$ so that $T_{\rm c}\sim
E_F\exp[-1/(k_FR_0)^2]$ is vanishingly small.  If the interaction strength is
tuned close to a Feshbach resonance \cite{Radzihovsky}, such that the
transition temperature strongly increases, then the particles have a
significant probability to be inside the centrifugal barrier at separations of
the order of $R_0$.  Under these conditions the system is very susceptible to
rapid losses arising from collisional relaxation into deep bound
states\cite{Gurarie,Castin}.

In contrast, for MW dressed molecules interacting via the potential $V_0(r)$,
the contribution $k_Fr^*$ to the attractive coupling strength provides a
significant transition temperature $T_{\rm c}$ even far from the resonance
associated with the presence of a two-molecule bound state. Given that the
molecules have a small probability to be at separations $\sim r_\delta$, in
this BCS regime one anticipates that the superfluid phase is not susceptible
to relaxation losses.

The dominant loss mechanism is from binary inelastic collisions
between $|+\rangle$ molecules, in which one or both are
transferred to the state $|-\rangle$ or $|1,-1\rangle$, which (since
$\delta > 0$) lie lower in energy than $|+\rangle$. For  $\Omega_R\alt\delta$, 
the released kinetic energy is $\sim\hbar\delta$ and can cause both molecules to escape from the
sample. The kinetic energy release requires a momentum transfer of
$\sim \hbar/\lambda_\delta$ with
$\lambda_\delta\equiv\sqrt{\hbar/M\delta}$.  For
${\lambda_\delta}/{r_\delta} \ll 1$ the particles cannot approach each
other sufficiently closely to allow the required momentum exchange,
and one anticipates a reduction in the loss rate. The same condition
can be derived semiclassically as the condition of adiabatic motion in
the potential. To go beyond this limit, and determine the
loss rate for $\lambda_\delta \sim r_\delta$, we have solved the full
two-body scattering problem, involving states of even parity which at
infinite separation are:
$(|+\rangle,|+\rangle),\,(|+\rangle,|-\rangle),\,(
+\rangle,|1,-1\rangle),\,(|-\rangle,|-\rangle),\,(|-\rangle,|1,-1\rangle)$
[the state $(|1,-1\rangle,|1,-1\rangle)$ is decoupled].  We calculate
(numerically) the probabilities $P_l$ that two $|+\rangle$-state
molecules with relative angular momentum $l$ are scattered into {\it
any} outgoing channel in which at least one of them is in the state
$|-\rangle$ or $|1,-1\rangle$. This corresponds to non-adiabatic
transitions from the potential $V_0(r)$ to the other potentials shown
in the inset to Fig.\ref{fig:bosurface}.\cite{2dcomment}

Taking into account that two molecules are lost in each inelastic collision,
and writing the molecule loss rate as $\dot n=-\alpha n^2$, for the 2D
inelastic rate constant we obtain:
$\alpha = 4{\hbar}/{M}\sum_{l}P_{l}$.
\begin{figure}[ttp]
\includegraphics[width=0.98\columnwidth]{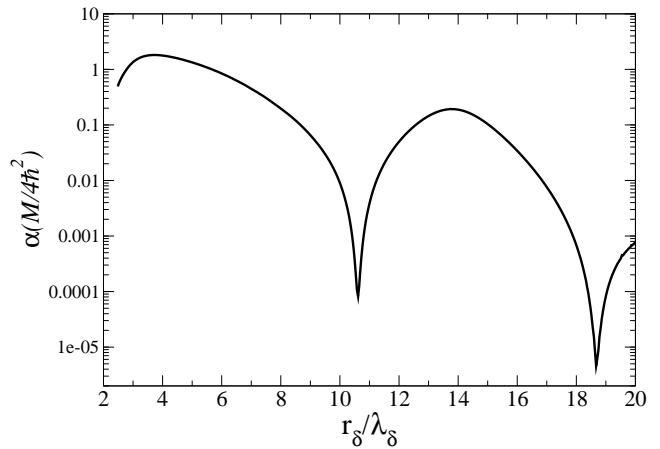}
\caption{ \label{fig:inelastic} Inelastic rate constant $\alpha$ as a function of
$r_\delta/\lambda_\delta$ for $\Omega_R = 0.25 \delta$ and $k r^*=1$ (see text).}
\end{figure}
For  incident energy chosen such that $kR^*\ll 1$, we have $\alpha\propto (kr^*)^2$.  
The results for this quantity versus $r_\delta/\lambda_\delta$
are shown in Fig.\ref{fig:inelastic}, for $\Omega/\delta=0.25$, and $k r^*
=1$.  The general trend is a reduction of inelastic losses with increasing
$r_\delta/\lambda_\delta$, consistent with the semiclassical expectations. However, in addition there is a dramatic modulation of
the inelastic scattering rate, arising from an interference of incoming and
outgoing waves in the scattering potential. By tuning to
$r_\delta/\lambda_\delta\simeq 10.5$, the rate constant can be suppressed to
$\alpha \simeq 4\times 10^{-4} \hbar/M$.

Thus, at a density $n= (10^8-10^9)\mbox{cm}^{-2}$ of, for example,
$^7$Li$^{40}$K molecules the lifetime of the gas is $\tau\approx
(\alpha n)^{-1} \simeq 2 - 0.2$ s.  The permanent dipole moment of
$^7$Li$^{40}$K in the ground state was found to be $3.5$ D
\cite{Dulieu}, and for $r_{\delta}/\lambda_{\delta}\simeq 10.5$ and
$\Omega_R=0.25\delta$ the lengthscales are $r_{\delta}\simeq 30$ nm,
$r^*\simeq 200$ nm. For $k_Fr^*$ close to unity we then
get $n\simeq 2\times 10^8$ cm$^{-2}$ and $E_F\simeq 120$ nK, so that
the transition temperature is $T_{\rm c}\simeq 10$ nK and
the lifetime is $\sim 1$ s.  (For $^{40}$K$^{87}$Rb, the possible $r^*$ 
is rather small and the high densities required for a sizeable 
$T_{\rm c}$ lead to rapid losses. The 
addition of a shallow optical lattice will increase the effective mass $M$, allowing $T_{\rm c}\sim 10$ nK at $n\sim 10^8$ cm$^{-2}$.)

We should avoid the presence of bound states of two molecules in the
potential $V_0(r)$, otherwise three-body recombination will lead
to a rapid decay of the gas on approach to the superfluid transition.  A
dimensional estimate for the three-body decay rate gives $\tau_{{\rm
    rec}}^{-1}\sim (\hbar r^{*2}/M)(k_Fr^*)^4n^2$, which can be large
for $k_F r^* \simeq 1$ and reasonable densities.  However, for $\Omega_R\simeq
0.25\delta$ (as used above) the potential $V_0(r)$ does not support bound
states for $r_{\delta}/\lambda_{\delta}\lesssim 14$. Thus, for the considered
value $r_{\delta}/\lambda_{\delta}\simeq 10.5$ the three-body recombination is
absent.

The formation of the $p_x+ip_y$ superfluid phase should be apparent in
numerous observables. These include quantities 
used to detect $s$-wave pairing in two component Fermi gases, such as the
density distribution, collective modes, and  RF absorption spectra \cite{Varenna,Trento}. 
The most striking new features of the $p_x+i p_y$ superfluid arise in
the presence of quantized vortices, which may be generated by rotation
of the gas as in usual superfluids.  RF absorption will then show
evidence for Majorana modes on the vortex cores
\cite{grosfeld:104516}. Ultimately one would hope to probe non-abelian
exchange statistics of these vortices \cite{tewari-2007-98}.

We would like to thank Jean Dalibard and Victor Gurarie for helpful
remarks. This work was supported by EPSRC Grant No. EP/F032773/1, by ANR Grant
06-NANO-014, by the IFRAF Institute, and by the Dutch Foundation FOM. LPTMS is
a mixed research unit No. 8626 of CNRS and Universit{\'e} Paris Sud.


\begin{thebibliography}{10}
\bibitem{gauge} The factors $\exp(-i\omega t)$ in
  Eqs.~(\ref{eq:g1}) and (\ref{eq:g}) arise from the time-dependence of the
  field, which has not been absorbed in the basis states.
\bibitem{comment1} For $\Gamma(\bm{k},\bm{q}) = V_0(\bm{k}-\bm{q})$, the
  interaction potential is rotationally symmetric, and the $p_x\pm ip_y$
  states are exactly degenerate. Owing to the circularly polarized MW, a
  (weak) splitting enters at higher order in $V_0$.
\bibitem{2dcomment} The outgoing particles are treated as 2D, which is valid
  both for $\lambda_\delta \gg l_z$ (when the energy release is
  insufficient to allow occupation of higher subbands) and for $\lambda_\delta
  \ll l_z$ (when the outgoing particles have a very small angle $\sim
  \lambda_\delta/l_z$ out of the 2D plane).
\end{thebibliography}

\end{document}